# EFFICIENT GAUSSIAN ELIMINATION ON A 2D SIMD ARRAY OF PROCESSORS WITHOUT COLUMN BROADCASTS

Mugurel Ionuț ANDREICA[1]

*În această lucrare se prezintă o metodă eficientă de implementare a eliminării Gaussiene pentru o matrice de dimensiuni n·m (m≥n), folosind o arhitectură SIMD ce constă dintr-o matrice de n·m procesoare. Algoritmul descris constă în 2·n-1=O(n) iterații, oferind astfel o creștere de viteză optimă față de versiunea serială. O particularitate a algoritmului este că nu necesită realizarea operației de broadcast decât pe liniile, nu și pe coloanele, matricii de procesoare. Lucrarea prezintă, de asemenea, o serie de extensii și aplicații ale algoritmului de eliminare Gaussiană.*

*This paper presents an efficient method for implementing the Gaussian elimination technique for an n·m (m≥n) matrix, using a 2D SIMD array of n·m processors. The described algorithm consists of 2·n-1=O(n) iterations, which provides an optimal speed-up over the serial version. A particularity of the algorithm is that it only requires broadcasts on the rows of the processor matrix and not on its columns. The paper also presents several extensions and applications of the Gaussian elimination algorithm.*

## 1. Introduction

The Gaussian elimination algorithm applied to an *n·m (m≥n)* matrix *A* consists of transforming the matrix into an equivalent upper triangular matrix *B* (i.e. *B(i,j)=0*, for *j<i*). The Gaussian elimination algorithm is a fundamental tool in a vast range of domains, like linear algebra (solving systems of linear equations [13], computing the rank of a matrix, computing the inverse of a matrix), geometry or scientific computing (molecular physics, geology, earthquakes). A serial version of the Gaussian elimination algorithm is presented below :

**SerialGauss(A,n):**
**for** *i=1* **to** *n-1* **do**
    // the search and swap stage
    **find** a **suitable** *(row,column) pair (r,c) (i≤r≤n, i≤c≤m), such that |A(r,c)|>0.*

---

[1] Assist., Computer Science and Engineering Department, Faculty of Automatic Control and Computer Science, Politehnica University of Bucharest, Bucharest, Romania, email: mugurel.andreica@cs.pub.ro

        **swap** *rows i and r in the matrix A*
        **swap** *columns i and c in the matrix A*
        // the reduction stage
        **for** *j=i+1* **to** *n* **do**
            *vaux=A(j,i)/A(i,i)*
            **for** *k=i* **to** *m* **do**
                *A(j,k)=A(j,k)-vaux·A(i,k)*

At the end of the algorithm, the matrix *A* is an upper triangular matrix. A *suitable* (row, column) pair *(r,c) (i≤r≤n, i≤c≤m)* is usually considered the one with the largest absolute value *|A(r,c)|*, because of numerical stability reasons. There are cases, however, when any (row,column) pair *(r,c)* is suitable, as long as *|A(r,c)|>0*. In such cases, we need to perform the swaps only if *A(i,i)=0*.

Because of its huge theoretical and practical importance, many parallel approaches for the Gaussian elimination technique were proposed. In [1, 2, 3], parallel Gaussian implementations using OpenMP and running on multiple processors (cores), as well as distributed implementations using MPI and running on multiple computing nodes were analyzed. Generating optimal task schedules for Gaussian elimination on MIMD machines was achieved in [4]. In [5], communication efficiency aspects of parallelizing the Gaussian elimination technique were considered. A systolic array implementation for dense matrices over GF(p) (the Galois field) was given in [6]. In [7], a Gaussian elimination algorithm over a synchronous architecture was presented, which is similar in several ways with the solution presented in this paper. Although the algorithm in [7] does not require any kind of broadcast mechanisms, it assumes that communication is possible between neighbouring processors located on the same row or column (at most *4* neighbours).

However, many of the mentioned parallel implementations seem to ignore the possibility that, at the $i^{th}$ iteration of the algorithm, the entry *A(i,i)* might be zero. This is because parallelizing the search and swap stage is more difficult. Such a hardness result was obtained in [8]. The time complexity of the serial algorithm is $O(n^2 \cdot m)$ (or $O(n^3)$ if *m=O(n)*) even if we do not search for a suitable entry *A(r,c)* at every iteration (and, thus, we perform no swaps). However, in the parallel case, the reduction stage is easy to parallelize, while the search and swap stage is not.

This paper proposes a novel approach for implementing the Gaussian elimination for an *n·m (m≥n)* matrix, using a SIMD 2D array of *n·m* processors (*n* rows with *m* processors each). The approach does not try to parallelize directly the serial algorithm, like many of the existing parallel solutions do; instead, it uses a different technique which allows matrix rows to *slide* past each other and reach their correct position in the final, upper triangular matrix. The solution does not search for the entry with the largest absolute value when setting the value *A(i,i)* and, because of this, it might not be numerically stable. However, it does permit row reorderings, due to the *sliding* mechanism, thus working under more realistic conditions than many other existing parallel solutions. Another particularity of the solution described in this paper is that it only requires mechanisms for broadcasting data on a row of processors and not on its columns, while several other parallel Gaussian elimination approaches require both row and column broadcasts. Since the broadcast requirements are reduced, the processor interconnection architecture can be simplified, thus reducing some of the architectural costs. Section 2 of this paper contains the description of the parallel algorithm and architecture, as well as a formal proof of correctness, based on induction over the number of rows of processors. Section 3 presents some validation results through simulations. In Section 4 several extensions and applications of the Gaussian elimination method are presented and in Section 5 we conclude and discuss future work.

## 2. Description of the Parallel Algorithm

The $n \cdot m$ processors are arranged into an $n \cdot m$ grid ($n$ rows with $m$ processors per row). Each processor *(i,j)* (located on row $i$ and column $j$) is connected to the processor on the row below it and on the same column. Processors on the $n^{th}$ row are connected to those on the first row. On each row $i$, processor *(i,i)* is a special processor and there are extra connections which allow processor *(i,i)* to broadcast data to all the other processors in its row. Each processor *(i,j)* has *6* registers: *tmp(i,j)*, *tmp2(i,j)*, *f(i,j)*, *cnt(i,j)*, *state(i,j)* and *state_changed(i,j)*. *tmp(\*,\*)* and *tmp2(\*,\*)* are used for temporary storage; *f(i,j)* is used for storing the final value of the entry on row $i$ and column $j$ of the triangular matrix obtained as a result of the algorithm. *cnt(i,j)* is a counter and is incremented after each iteration of the algorithm (*cnt(i,j)* is identical for all the processors and could be implemented as a single shared register, instead of $n \cdot m$ distributed registers). *state(i,j)* stores the state of the processor. If (*state(i,j)=1*), then the value stored in *f(i,j)* is the final value of the *(i,j)* entry of the resulting triangular matrix; otherwise (if *state(i,j)=0*), *f(i,j)* contains no meaningful value. The states of all the processors on the same row $i$ are identical; thus, a single shared register for each row could be used. *state_changed(i,j)* is a boolean register which holds the value *true* if the state of the processor changed during the current iteration. Except for these registers, each processor stores its row and column (*i* and *j*) in two special read-only registers.

In the beginning of the algorithm, entry *(i,j)* of the initial matrix is stored in *tmp(i,j)*, *cnt(\*,\*)* and *state(\*,\*)* are set to *0* and *state_changed(\*,\*)* are set to *false*. We will perform *2·n-1* iterations. At each iteration, the *tmp* value of each processor *(i,j)* is transferred to the *tmp* value of the processor *(i+1,j)* (processor *(n,j)* transfers the data to processor *(1,j)*). All the transfers occur simultaneously, as in any SIMD computer. After the data transfer, each processor performs a series of computations, described by the function *Compute* (see below). The algorithm works as follows. At each iteration $t$ ($1 \leq t \leq 2 \cdot n-1$), only the processors on rows $i \leq t$ perform meaningful computations (this is enforced by the test $cnt(i,j) \geq i$ in the pseudocode below). If the state of the processors on row $i$ is *0*, then we verify if the current row stored in the $tmp(i,*)$ values could be the final row to be stored on row $i$ of the resulting upper triangular matrix. Processor $(i,i)$ performs this test, by comparing $|tmp(i,i)|$ against *0*. If $(|tmp(i,i)|>0)$, then we can store the current $tmp(i,*)$ values into the $f(i,*)$ values, because the condition that the entry $(i,i)$ is non-zero is fulfilled. The processor $(i,i)$ broadcasts a *changed state announcement* to all the other processors on row $i$. If this announcement is *1* $(|tmp(i,i)|>0)$, then the new state of the processors $(i,j)$ becomes *1* and the $tmp(i,*)$ values are copied into the corresponding $f(i,*)$ values; after this, $tmp(i,*)$ is set to *0*; if the announcement is *0*, then no more computations are performed by the processors on row $i$ during the current iteration.

**Compute(processor (i,j)):**
*cnt(i,j)=cnt(i,j)+1*
**if** *(cnt(i,j)≥i)* **then**
    **if** *(state(i,j)=1)* **then**
        **if** *(i=j)* **then**
            *tmp2(i,i)=tmp(i,i)/f(i,i)*
            **broadcast** *tmp2(i,i) to all the processors (i,j) on row i, with i≠j*
        **else** // i≠j
            *tmp2(i,j)=the value broadcasted by processor (i,i)*
        *tmp(i,j)=tmp(i,j)-tmp2(i,j)·f(i,j)*
    **else** // (state(i,j)=0)
        **if** *(i=j)* **then**
            **if** *(|tmp(i,i)|>0)* **then**
                **broadcast** *true to all the processors (i,j) on row i, with i≠j*
                *state_changed(i,i)=true*
            **else**
                **broadcast** *false to all the processors (i,j) on row i, with i≠j*
        **else** // i≠j
            *state_changed(i,j)=the value broadcasted by processor (i,i)*
        **if** *(state_changed(i,j)=true)* **then**
            *state(i,j)=1*
            *state_changed(i,j)=false*
            *f(i,j)=tmp(i,j)*
            *tmp(i,j)=0*

If, at the beginning of the iteration, the state of the processors on row $i$ is *1*, then the values stored in $tmp(i,*)$ are processed accordingly. The value $tmp2(i,i)=tmp(i,i)/f(i,i)$ is computed by processor $(i,i)$ and then broadcasted to all the other processors on row $i$ (which store it in their own *tmp2* register). Afterwards, each value $tmp(i,j)$ ($1 \leq j \leq m$) is decreased by $tmp2(i,j) \cdot f(i,j)$. As a result of this, $tmp(i,i)$ becomes *0*. At the next iteration, the processors on the row below row $i$ receive the decremented $tmp(i,*)$ values of the processors on row $i$.

We will prove the following statement: **the $tmp(i,j)$ values received by a row $i$ of processors at an iteration $t \geq i$ are zero for $1 \leq j \leq i-1$**. The proof is based on induction over the row $i$ of processors. The base case consists of $i=1$ where it is obvious that the statement holds. Let's assume that the statement holds for the first $i-1$ rows of processors and we will now try to prove it for the $i^{th}$ row, too.

Let's analyze all the possible cases. If, at the beginning of the previous iteration, $state(i-1,*)=1$, then according to the algorithm, the entry $tmp(i-1,i-1)$ was decremented to $0$. The entries $tmp(i-1,j)$ $(1 \leq j \leq i-2)$ were zero and they were further decremented by $tmp2(i-1,j) \cdot f(i-1,j)$. However, the statement is valid for row $i-1$ and any previous iteration $t \geq i-1$ and, thus, also for the iteration $t'$ when the state of row $i-1$ changed from $0$ to $1$. At iteration $t'$, the values $f(i-1,j)$ $(1 \leq j \leq i-2)$ were set to $tmp(i-1,j)$, which were equal to $0$. Thus, we have $f(i-1,j)=0$ $(1 \leq j \leq i-2)$ and, as a consequence, the $tmp(i-1,j)$ $(1 \leq j \leq i-2)$ values were, in fact, not decremented at all. So, at the end of the previous iteration, all the $tmp(i-1,j)$ $(1 \leq j \leq i-1)$ values were $0$, after which they were transferred to the row $i$ of processors. We also have to analyze the case where $state(i-1,*)=0$ at the beginning of the previous iteration. If, in the previous iteration, the row of processors $i-1$ changed its state from $0$ to $1$, then it set all the $tmp(i-1,*)$ values to $0$ and these were then received by the next row of processors; thus, the statement holds. The only case left is when, at the beginning of the previous iteration, $state(i-1,*)=0$ and it does not change to $1$. In this case, the $tmp(i-1,*)$ values are not changed – they are transferred at row $i$ as they are. Since the statement holds for the rows up to $i-1$, we have $tmp(i-1,j)=0$ $(1 \leq j \leq i-2)$. However, because the state of the previous row of processors did not change to $1$, we must also have $tmp(i-1,i-1)=0$. This concludes the proof.

Using the statement proved above, it is easy to conclude that the final values $f(i,j)$ are $0$ $(1 \leq j \leq i-1)$ and, thus, the obtained matrix is upper-triangular. We only have one problem left. It is possible that, after $2 \cdot n-1$ iterations, some rows of processors are still in state $0$. If a row $i$ of processors is still in state $0$, then none of the remaining matrix rows have non-zero values in the $i^{th}$ column (and, thus, the initial matrix is singular). Considering that every remaining matrix row also has a zero entry in every column $j$, where row $j$ of processors is in state $1$, this means that every remaining matrix row is full of zeroes. We have two choices. We can either report that the determinant is zero (in case it should have been non-zero), or we can set $f(i,j)=0$ for all the processors with $state(i,j)=0$ after the $2 \cdot n-1$ iterations. From the point of view of a SIMD implementation, at the beginning of each iteration, after performing the data transfers, we select all the processors and let them increment the counter $cnt(i,j)$. Then, we select all the processors $(i,i)$ with $state(i,i)=1$ and $cnt(i,i) \geq i$ and let them compute the value $tmp2(i,i)$ and broadcast it to the other processors in their row. Afterwards, we select all the

Mugurel Ionuţ Andreica

processors *(i,j)* with *i≠j*, *state(i,j)=1* and *cnt(i,j)≥i* and let them receive the value broadcasted by the processor *(i,i)* on their row. After this, we select all the processors *(i,j)* with *state(i,j)=1* and *cnt(i,j)≥i* and let them modify the value *tmp(i,j)* accordingly. In the second part of the iteration we select all the processors *(i,i)* with *state(i,i)=0*, *|tmp(i,i)|>0* and *cnt(i,i)≥i* and let them perform the broadcast of the *changed state announcement* (*1*) to their row of processors and the assignment *state_changed(i,i)=true*. Afterwards we select all the processors *(i,i)* with *state(i,i)=0*, *|tmp(i,i)|=0* and *cnt(i,i)≥i* and let them perform the broadcast of the *changed state announcement* (*0*). We then select all the processors *(i,j)* with *i≠j*, *state(i,j)=0* and *cnt(i,j)≥i* and let them receive the *changed state announcement* broadcasted by the processor *(i,i)* on their row. After this, we select all the processors *(i,j)* with *state_changed(i,j)=true* and *cnt(i,j)≥i* and we let them perform the assignments *state(i,j)=1*, *f(i,j)=tmp(i,j)*, the clearance to *0* of *tmp(i,j)* and the clearance to *false* of the register *state_changed(i,j)*.

A sample architecture on which the previously described algorithm can run is depicted in Fig. 1.

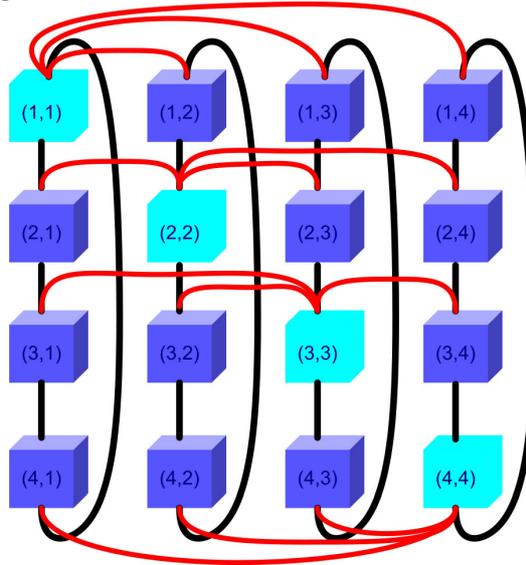

Fig. 1. A sample SIMD 2D array of processors with n=m=4.

### 3. Validation through Simulations

The described parallel algorithm was implemented using the Parallaxis parallel programming language [9]. A suite of tests were performed, which were meant to verify the correctness of the parallel implementation. Because of this, a serial Gaussian elimination algorithm was also implemented and both algorithms were executed on *20 n·(n+1)* randomly generated matrices, with *n* ranging from *1* to *50*. Since the parallel and serial implementations are based on slightly different sequences of actions, the outputs of the two algorithms were not identical. However, by interpreting the two output matrices as the (augmented) matrices of two systems of linear equations and solving these systems, we obtained two sets of values of the *n* variables. After sorting the sets, both algorithms should have obtained the same sequence of values of the variables. During testing, any singular matrix that occured was discarded and another matrix was regenerated instead. Furthermore, the determinants of the upper-triangular matrices obtained by both algorithms were also computed (the determinant is equal to the product of the elements on the main diagonal of a lower- or upper-triangular matrix) and their absolute values were compared against each other (because of row reorderings, the sign of the determinant may change, but not its absolute value). Since the obtained matrices may have more than *n* columns, the determinant was computed considering only the first *n* columns. In order to accurately compute the determinant, a library for large real numbers was used [10] (because the determinant's value exceeded the standard *double* or *long double* types). For each of the *20* matrices, all the tests were *passed* : the same determinants (in absolute value) and the same values of the variables were obtained.

### 4. Extensions and Applications of the Gaussian Elimination Method

The function ***Compute***, presented in Section 2, can be extended, for instance, by considering that all the values of the initial matrix are integers and that all the operations (addition, subtraction, multiplication, division) are performed *modulo* a (prime) number *M*. It is easy to perform addition, subtraction and multiplication modulo *M* (we first compute the result *R* normally, and then take its remainder, i.e. *R mod M*). The division operation requires the existence of a *multiplicative inverse*. The multiplicative inverse of a number *x* is the number $x^{-1}$ such that $x \cdot x^{-1} = 1 \pmod{M}$. $x^{-1}$ can be computed by using the extended Euclidean algorithm [11]. A somewhat easier case occurs when *M=2*. In this case, addition and subtraction are equivalent to *xor*, multiplication is equivalent to *and*, and division is easy (since we only divide a number *x* by *1*, the result is always *x*).

A matrix operation which is related to the Gaussian elimination operation is matrix multiplication. Multiplying two *k*-by-*k* (square) matrices can be easily performed in $O(k^3)$ time or, if we use Strassen's algorithm [12], in $O(k^{2.807})$ time. Matrix multiplication has applications in many domains. We will consider next an application to Combinatorics. Let's consider the set of sequences whose elements belong to the set *{1,...,k}*, for which we are given a binary *k*-by-*k* "transition" matrix *T(i,j)*. If *T(i,j)=1*, then the element *j* can be located right after element *i* in the sequence (if *T(i,j)=0*, it cannot). We want to compute the number of (valid) sequences with *n* elements (mod *M*). The straight-forward dynamic programming solution is the following. We compute *S(l,j)*=the number of (valid) sequences with $l$ elements whose last element is $j$ ($1 \leq j \leq k$). We have $S(1,j)=1$. For $l>1$, we have $S(l,j)$=the sum of the values $(T(i,j) \cdot S(l-1,i))$ ($1 \leq i \leq k$). The answer is the sum of the values $S(n,i)$ ($1 \leq i \leq k$). The time complexity is $O(n \cdot k^2)$. In order to improve the time complexity (to $O(k^3 \cdot log(n)+k^2)$), we will consider the $k$-element column vectors $SC(l)$, where $SC(l)(j)=S(l,j)$. We have $SC(l)=T \cdot SC(l-1)$. Thus, $SC(n)=T^{n-1} \cdot SC(1)$. By efficiently raising the transition matrix $T$ at the $(n-1)$-th power (e.g. by using repeated squaring), we obtain an efficient method of computing the column vector $SC(n)$ and the answer (the sum of the values $SC(n)(i)$, with $1 \leq i \leq k$). A second efficient method ($O(k^4 \cdot log(n))$) to compute the number of valid sequences of length $n$ is to compute the values $S(i,l,j)$=the number of sequences of length $2^l$, for which the first element is $i$ and the last element is $j$. We have $S(i,0,i)=1$ and $S(i,0,j \neq i)=0$. For $l>0$, $S(i,l,j)$=the sum of the values $S(i,l-1,p) \cdot T(p,q) \cdot S(q,l-1,j)$, with $1 \leq p \leq k$ and $1 \leq q \leq k$. Afterwards, we write the number $n$ as a sum of powers of $2$, i.e. $n=2^{pow(0)}+\ldots+2^{pow(B)}$. We will compute the values $U(i,l,j)$=the number of sequences starting with $i$, ending with $j$, and whose length is $2^{pow(0)}+\ldots+2^{pow(l)}$. Obviously, we have $U(i,0,j)=S(i,pow(0),j)$. For $l \geq 1$, we have $U(i,l,j)$=the sum of the values $U(i,l-1,p) \cdot T(p,q) \cdot S(q,pow(l),j)$, with $1 \leq p \leq k$ and $1 \leq q \leq k$. The answer is the sum of the values $U(i,B,j)$ ($1 \leq i \leq k$ and $1 \leq j \leq k$). In both methods, all the arithmetic operations are perfomed modulo $M$.

    An interesting application of the Gaussian elimination method is the following. We consider $N$ numbers $A(i)$ ($0 \leq A(i) \leq 2^B-1$, $1 \leq i \leq N$). We want to find a subset $\{i_1, \ldots, i_k\}$ of $\{1, \ldots, N\}$, such that $A(i_1)$ xor $A(i_2)$ xor $\ldots$ xor $A(i_k)$ is maximum ($k$ may be any number). Let's denote this maximum value by $XM$. We will compute $XM$ bit by bit (from the bit $B-1$, the most significant one, down to the bit $0$, the least significant one). Let's assume that we computed the values $BV(B-1), BV(B-2), \ldots, BV(i+1)$ of the bits $B-1, \ldots, i+1$ of $XM$ and we now want to compute $BV(i)$. In order to achieve this, we will use the following transformation. We will obtain a system of equations in base $2$, having $B-i$ equations and $N$ unknown variables. The $N$ unknown variables, $x(1), \ldots, x(N)$ can be either $0$ or $1$. If $x(j)=1$ ($1 \leq j \leq N$), then $i$ belongs to the subset we want to compute; otherwise, $i$ does not belong to this subset. The coefficients of this system, $c(p,q)$ ($i \leq p \leq B-1$, $1 \leq q \leq N$), are: $c(p,q)$=the $p^{th}$ bit of $A(q)$. This way, equation $p$ has the following structure: $c(p,1) \cdot x(1)$ xor $c(p,2) \cdot x(2)$ xor $\ldots$ xor $c(p,N) \cdot x(N)=c(p, N+1)$, where: $c(p, N+1)=BV(p)$ ($i+1 \leq p \leq B-1$) and $c(p, i)=1$. We will run the Gaussian elimination method on the extended matrix of the system of linear equations (which has $B-i$ rows and $N+1$ columns), where all the operations are performed modulo $2$ (addition and subtraction are equivalent with *xor*, multiplication is equivalent with *and*, and division is not necessary). However, we will never consider column $N+1$ as a candidate for swapping with another column. After performing the Gaussian elimination, the extended matrix of the system ($c(*,*)$) will have $1$ on the main diagonal from row $i$ up to a row $q$ ($i-1 \leq q \leq B-1$) (rows are numbered from $i$ to $B-1$)

and $0$ on the rows $q+1, ..., B-1$. Under the main diagonal, the matrix will contain only $0$ elements. The system of linear equations has a solution if we have $c(j,N+1)=0$ on all the rows $j=q+1, ..., B-1$. If the system has a solution, then $BV(i)=1$, otherwise $BV(i)=0$. This way, we can compute *XM* bit by bit. The subset of indices $\{i_1, ..., i_k\}$ is computed from the solution of the last system of equations. The time complexity of this algorithm is $O(B^3 \cdot N)$. We can reduce the time complexity to $O(B^2 \cdot N)$, as follows. We notice that the system of equations corresponding to the bit $i$ has the first $B-i-1$ rows identical to those of the system of equations corresponding to the previous bit $(i+1)$. Thus, we will keep the matrix obtained as a result of the Gaussian elimination performed at the bit $i+1$, to which we add a new row, corresponding to the bit $i$ (the $j^{th}$ value of this row is equal to the $i^{th}$ bit corresponding to $A(o(j))$, where $o(j)$ is the index corresponding to column $j$, $1 \leq j \leq N$; initially, $o(j)=j$, but we have to swap $o(j)$ and $o(k)$ whenever we swap the columns $j$ and $k$ between them; the $(N+1)^{st}$ value of the row is $1$). Let $r(i+1)$ be the row corresponding to the row added when considering the bit $i+1$. If $BV(i+1)=0$ then we will set $c(r(i+1),N+1)=not\ c(r(i+1),N+1)$ (i.e. we change its value into the opposite one). We will also store the first row $q(i+1)$ where $c(q(i+1),q(i+1))=0$ (i.e. there were no more $1$ elements on row $q(i+1)$ or below it). We will reduce the newly added row by subtracting from it all the rows having a $1$ on their main diagonal position. Then, if this row contains any $1$ elements, we will swap it with the row $q(i+1)$; afterwards, we will swap the column $q(i+1)$ with a column $C$ containing a $1$ element on the reduced newly added row and we will set $q(i)=q(i+1)+1$. If, after being reduced, the newly added row contains no $1$ elements, we set $q(i)=q(i+1)$. Note that this time we considered that the rows were numbered from $1$ to $B-i$. Checking if $BV(i)=1$ can be performed in $O(B)$ time (by considering every row $h$ from $q(i)+1$ to $B-i$ and checking that $c(h,N+1)=0$). We will set $r(i)$ to the index of the matrix row on which the newly added row is located. By using this improvement we basically perform only one Gaussian elimination over the whole course of the algorithm. Thus, the time complexity becomes only $O(B^2 \cdot N)$. Fig. 2 depicts the extended matrix on which the Gaussian elimination is performed.

   A problem which is very similar to the previous one, and yet it has a totally different solution is the following. We consider a sequence of $N$ natural numbers $A(i)$ ($0 \leq A(i) \leq 2^B-1$, $1 \leq i \leq N$). We want to find a contiguous subsequence $A(i), ..., A(j)$ ($A(i), ..., A(j)$ are located on consecutive positions in the sequence), such that $A(i)\ xor\ A(i+1)\ xor\ ...\ xor\ A(j)$ is maximum. We will compute the *prefix xors* (similar to the well-known *prefix sums*): $X(0)=0$ and $X(1 \leq i \leq N)=X(i-1)\ xor\ A(i)$. Then, we will maintain a trie (a prefix tree), in which we will introduce, one step at a time, these prefix xors. Each prefix xor will be interpreted as a binary string with $B$ elements. The first element of the string will be the most significant bit (the bit $B-1$), which will be followed by the second most significant bit, and so on, until the least significant bit (the bit *0*). Initially, we will introduce in the trie the string corresponding to *X(0)*. We will traverse the sequence from *1* to *N*, and for each position *i*, we will compute the largest xor of a contiguous subsequence ending at position *i*. Let's consider *BV(i, B-1), ..., BV(i, 0)* to be the bits of the largest xor of a contiguous subsequence ending at position *i*. We will traverse the binary string corresponding to *A(i)* (which consists of *B* bits) with an index *j*, starting from the bit *B-1*, down to the bit *0*. We will denote by *A(i,j)* the $j^{th}$ bit of *A(i)*. At the same time, we will maintain a pointer *pnod* to a node of the trie, which will be initialized to the trie's root. At every bit *j*, we verify if *pnod* has an edge labelled with *(1-A(i,j))* towards one of his sons. If it does, then *BV(i,j)=1* and we set *pnod* to the son corresponding to the edge labelled with *(1-A(i,j))*. If *pnod* does not have such an edge, then *BV(i,j)=0* and we set *pnod* to the son corresponding to the edge labelled with *A(i,j)*. After computing *BV(i,*)* we will insert the string corresponding to *X(i)* in the trie. The result will be the largest value among those corresponding to the strings *BV(i,*)* (interpreted as numbers having *B* bits). The time complexity of this algorithm is *O(N·B)*.

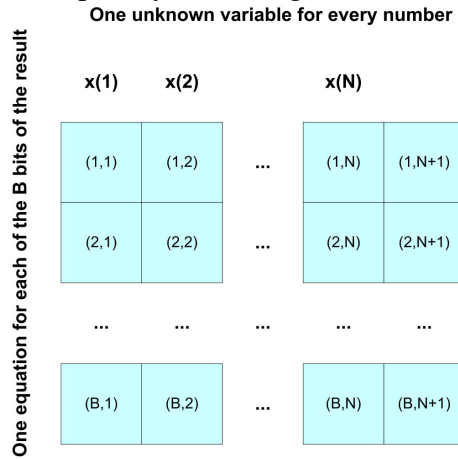

Fig. 2. The matrix on which the Gaussian elimination is performed.

The problem can be extended by considering the following constraints: the length of the computed sequence must be at least *L* and at most *U* (*1≤L≤U≤N*). In this case, we will have to remove some prefix xors from the trie. To be more exact, we will proceed like in the previous algorithm. When we reach the position *i* and we want to compute *BV(i,*)*, we perform the following action: if *i>U*, then we remove from the trie the string corresponding to *X(i-U-1)*. We will compute *BV(i,*)* only if *i≥L*; after computing *BV(i,*)* (*i≥L*), we will insert into the trie the string corresponding to *X(i-L+1)*. In order to be able to remove strings from a trie, we will store for each trie node *pnod* the number of strings which contain this node (when we add a new string, we increment by *1* the counters of all the nodes encountered along the path from the root, and when we remove a string, we decrement by *1* the counters of the same nodes visited when the node was inserted). If, at some point, the counter of a node *pnod* (different from the trie's root) becomes 0, we can remove the edge (labelled with *E*) between *pnod* and its parent (this way, *pnod*'s parent will not have *pnod* as a son or an edge labelled with *E* anymore). The time complexity in this case stays the same (i.e. *O(N·B)*). We notice that, although this problem and the previous one were apparently very similar, the algorithmic techniques required for obtaining a solution are quite different (we didn't use any Gaussian elimination for this problem).

Another application of the Gaussian elimination method is the following. We consider an undirected graph with *N* vertices. There is a light bulb in every vertex. The bulb in vertex *i* is initially in a state *SI(i)* (*SI(i)=1* means *"on"*, and *SI(i)=0* means *"off"*). Every vertex *i* also has a cost $C(i) \geq 0$ which needs to be paid if we want to modify its state (from *on* to *off*, or from *off* to *on*). We want to bring every bulb *i* into a final state *SF(i)* (which may be identical to *SI(i)*), by performing a sequence of the following type of actions: we *touch* the bulb *i* (and we pay the cost *C(i)*) – as a consequence of this action, the state of the bulb *i* and of all the neighbouring bulbs are changed (but we do not have to pay anything for the neighbouring bulbs). We want to find a strategy which brings the bulbs to their final states, such that the total cost is minimum. The general case can be solved as follows. We notice that we never have to touch the same bulb twice. We will associate a variable *x(i)* to every bulb *i*, which will be either *0* or *1*, representing the number of times the bulb *i* was touched. We will construct a system of linear equations in base *2*. We will have an equation for each bulb *i* ($1 \leq i \leq N$): *c(i,1)·x(1) xor c(i,2)·x(2) xor ... xor c(i,N)·x(N)=c(i,N+1)*. The coefficients *c(i,j)* will be *1* for *j=i* and for those bulbs *j* for which the edge *(i,j)* exists in the graph; for the other bulbs *j*, the coefficients *c(i,j)* will be *0*; *c(i,N+1)=SI(i) xor SF(i)*. We will use the Gaussian elimination technique on this system of equations, where all the operations will be performed modulo *2*. Afterwards, the extended matrix *c(\*,\*)* of the system will have *1* elements on the main diagonal on the first *PR* rows (*PR* is obtained after running the elimination method), and the last *N-PR* rows will have *c(i,j)=0* ($PR+1 \leq i \leq N$, $1 \leq j \leq N$). If we have *c(i,N+1)=1* (for some $PR+1 \leq i \leq N$), then the problem has no solution. Otherwise we have *N-PR* free variables – the ones corresponding to the last *N-PR* columns. We must pay attention that during the algorithm, the columns may be swapped among each other, such that the last *N-PR* columns do not necessarily correspond to the initial variables *x(PR+1), ..., x(N)*. For the *N-PR* free variables we will have to try every possible combination of assigning values to them (there are $2^{N-PR}$ possibilities overall). For every combination we will compute the values of the other *PR* variables (the *bound* variables). Once we compute the values of all the variables *x(1), ..., x(N)*, the cost of the strategy is *CS=C(1)·x(1)+...+C(N)·x(N)*. We will choose the strategy with minimum cost among all the $2^{N-PR}$ possibilities.

This method could lead to a significant improvement upon a naïve algorithm which tries each of the $2^N$ possible combinations of assigning values to the *N* variables. However, the degree of improvement depends on the structure of the graph. For instance, let's assume that the graph consists of *P·Q* vertices, arranged on *P* rows and *Q* columns. Each vertex *(i,j)* (on row *i*, column *j*) is adjacent to the vertices to the north, south, east and west (if these neighbours exist). In this case, we can consider the $2^Q$ (or $2^P$) possibilities of assigning values to the variables of the vertices on the first row (first column). Let's consider the *first row* case (the first column case is handled similarly). After assigning values to the variables of the vertices on the first row, we will traverse the other vertices in increasing row order (with *i* from row *2* to *P*) and, for equal rows, in increasing column order (with *j* from *1* to *Q*). The value of the variable *x(i,j)* is uniquely determined by the values of the variables computed earlier. If the vertex *(i-1,j)* is in its final state (considering the values *x(\*,\*)* of itself and of all of its neighbours except for *(i,j)*), then *x(i,j)=0*; otherwise, *x(i,j)=1*. If the vertices on the last row are not in their final states (considering the values *x(\*,\*)* of themselves and of all of their neighbours), then the assignment of values to the variables of the vertices on the first row did not lead to a solution. Otherwise, we obtained a solution and we compute its cost (like before). Fig. 3 shows the structure of the *P*x*Q* graph.

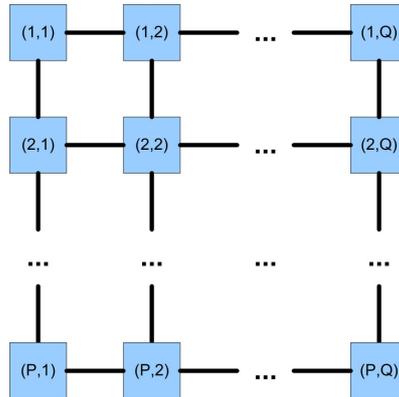

Fig. 3. The structure of the *P*x*Q* graph in the first case.

Let's consider now the same graph, where every vertex is adjacent to its (at most) *4* diagonal neighbours. We will color every vertex *(i,j)* in white, if *(i+j)* is even, or in black, if *(i+j)* is odd. This way, we obtained (at most) *2* connected components, such that all the vertices of the same component have the same color. For every component, we will split the vertices *(i,j)* in groups, according to the values *min{i,j}* (all the vertices *(i,j)* with the same value of *min{i,j}* will belong to the same group). We will consider every possibility of assigning values to the variables of the vertices *(i,j)* for which *min{i,j}* is minimum among all the vertices of the component. For each possibility, we will consider all the other groups in the component, in increasing order of *min{i,j}*. For each such group, we will traverse the vertices *(i,j)* for which *j≤i* from the lower-indexed row to the higher-indexed row, along the column *j*. Then we will traverse the vertices *(i,j)* for which *i≤j* from the lower-indexed column to the higher-indexed column, along the row *i*. Every time we reach a vertex *(i,j)*, this vertex will have a neighbour *(i',j')* for which *min{i',j'}=min{i,j}-1* and for which the vertex *(i,j)* is the last vertex whose variable has not been assigned a value, yet. Thus, the variable *x(i,j)* will be uniquely determined, based on the current state of *(i',j')* and its desired final state (*0*, if the two states are the same, and *1* otherwise). Like in the previous case, some initial assignments may not lead to a solution (if some vertices do not end up in their desired final states).

One last case we consider here is that in which every vertex *(i,j)* is adjacent to all of its (at most) *8* neighbours (on the same row, column, or diagonal). In this case, we will assign values to the variables of the vertices on the first row and column. For each assignment, we will traverse the vertices *(i,j)* like in the previous case, according to *min{i,j}*. Basically, we traverse the vertices on column *2* (from lower-indexed rows to higher-indexed rows), followed by the vertices on row *2* (from lower-indexed columns to higher-indexed columns), and so on. Like in the previous case, when we reach a vertex *(i,j)* during the traversal, this vertex will have at least one neighbour *(i',j')* (with its variable assigned to some value), for which *(i,j)* is the only neighbour whose value has not been assigned, yet. Thus, *x(i,j)* will be uniquely determined. In this case, we have $2^{P+Q-1}$ possibilities for assigning variables to the vertices *(i,j)* on the first row and column (those with *min{i,j}=1*).

A problem related to the previous ones is the following. We consider a matrix with *M* rows and *N* columns. Each cell *(i,j)* of the matrix contains a bulb. The bulb *(i,j)* is in the initial state *SI(i,j)* (*1-on* or *0-off*) and must be brought into its final state *SF(i,j)*. The only operations we can perform are: changing the state of all the bulbs on a row *i*, which costs *CL(i)≥0* (*1≤i≤M*), and changing the state of all the bulbs on a column *j*, which costs *CC(j)≥0* (*1≤j≤N*). We want to find a sequence of operations with minimum total cost such that, in the end, every bulb is in its final state. Like before, we notice that we never need to perform an operation on a given row or column more than once. We will denote by *xL(i)* (*xC(j)*) the variables which describe if we perform (*1*) or not (*0*) an operation on the row *i* (column *j*). We will consider two cases. In the first case, we consider that we do not perform an operation on row *1*: *xL(1)=0*. We will traverse all the bulbs *(1,j)* on the row *1* and, if *SI(1,j)=SF(1,j)*, then we must not perform an operation on column *j* (*xC(j)=0*); otherwise, we must perform an operation on column *j* (*xC(j)=1*). Then, we traverse all the bulbs *(i,1)* on the column *1* (*i≥2*): if *SI(i,1) xor xC(1)=SF(i,1)*, then *xL(i)=0*; otherwise, *xL(i)=1*. Then we verify, for every bulb *(i,j)*, if *SI(i,j) xor xL(i) xor xC(j)=SF(i,j)*. If the condition is *true* for every bulb, then we obtained a solution and we compute its cost: *CS=* $CL(1) \cdot xL(1) + ... + CL(M) \cdot xL(M) + CC(1) \cdot xC(1) + ... + CC(N) \cdot xC(N)$. In the second case we will consider *xL(1)=1*. In a similar manner we will compute the values *xC(j)*: if *SI(1,j) xor xL(1)=SF(1,j)*, then *xC(j)=0*; otherwise, *xC(j)=1*. Then we compute the values *xL(i)* (*i≥2*): if *SI(i,1) xor xC(1)=SF(i,1)*, then *xL(i)=0*; otherwise, *xL(i)=1*. As before, we verify if we obtained a solution. The minimum cost is given by the minimum of the (at most) *2* possible solutions.

A strongly related problem is the following. We have the same input data and objective, but we have a different set of operations that can be performed. An operation consists of choosing a bulb *(i,j)* and changing the state of all the bulbs on the row *i* and the column *j* (except for the bulb *(i,j)*). The cost of such an operation is *1* – thus, we want to minimize the total number of operations required for bringing every bulb in its final state. An operation in this problem is equivalent to changing the state of a row, followed by that of a column (in the terms of the previous problem). At first, we will consider that we can perform the same operations like in the previous problem. Thus, using that solution, we obtain (at most) two distinct solutions. The solution *i* (*i=1,2*) consists of changing the states of the bulbs on the rows *L(i,1), ..., L(i,nL(i))* and on the columns *C(i,1), ..., C(i,nC(i))*. We will try to transform this solution into one which is compatible with the operations allowed in this problem. We will construct pairs of the form *(row, column)*: *(L(i,1), C(i,1)), ..., (L(i,K), C(i,K))*, where *K=min{nL(i), nC(i)}*. A pair *(L(i,j), C(i,j))* means that the bulb *(L(i,j), C(i,j))* is chosen for an operation. If *nL(i)>K*, then the extra number of rows *(nL(i)-K)* must be even. We will form the pairs *(L(i,K+1), 1), (L(i,K+2), 1), ..., (L(i,nL(i)), 1)* (i.e. we choose every bulb *(L(i,K+1), 1), ..., (L(i,nL(i)), 1)* in an operation). If *nC(i)>K*, then *(nC(i)-K)* must be even and we will choose the bulbs *(1, C(i,K+1)), ..., (1,C(i,nC(i)))* in an operation. We will choose that solution *i* (*i=1,2*) for which we obtain the minimum number of operations.

The problem of turning light bulbs on and off on a graph (where, when touching a bulb, the states of the bulb and of all of its neighbours change), can be solved without using the Gaussian elimination technique on graphs with special structures, particularly graphs with a tree-like structure, like trees, trees of cycles or graphs with bounded treewidth (when a suitable tree decomposition is also given). In order to make an informed choice, we first need to recognize the class to which the input graph belongs. Thus, graph recognition algorithms are important in this situation. We will present next a simple graph recognition algorithm for the class of trees of cycles.

A *tree of cycles* is an undirected graph with one of the following attributes:
- it is a cycle of any length *K* (*K≥3*)
- it is a graph obtained by attaching a cycle *C* of some length *K* (*K≥3*) to an edge of a tree of cycles *CT*

Attaching a cycle to an edge *e* of a graph means replacing an edge of the cycle with the edge *e* of the graph and replacing the two vertices of the cycle connected by the replaced edge by the two vertices of the graph connected by *e*. Fig. 4 depicts the operation of attaching a cycle to a tree of cycles.

In order to decide if a given undirected graph is a tree of cycles, we will perform the following operation repeatedly (for vertex *i*): if there exists a vertex *i* with degree *2*, and its neighbours are *j* and *k*, then we remove the vertex *i* from the graph (and its two adjacent edges) and we introduce instead the edge *j-k*, if this edge does not exist in the graph already. If the given graph is a tree of cycles then, in the end, we will only have two vertices left (initially, the number of vertices of the graph must be at least *3*). In order to perform the described operation, we will use a queue in which we will insert all the vertices which initially have degree *2*. Then, we will repeatedly extract vertices from the queue (as long as the queue is not empty). Every time we extract a vertex *i* from the queue we perform the operation for *i*. Thus, there exists the possibility of modifying the degrees of its two neighbours *j* and *k*. If, after performing the operation for the vertex *i*, the vertex *j* (*k*) ends up with degree *2* (before the operation it had a degree greater than *2*), we insert *j* (*k*) at the end of the queue. We now only need to use an efficient data structure for checking quickly if two vertices are adjacent in the graph, for finding the two neighbours of a vertex *i*, and for removing edges from the graph. We can use, for instance, balanced trees, in which we store *2* pairs *(i,j)* and *(j,i)* for each edge *i-j* which (currently) exists in the graph. This way, we can check easily if a given edge exists in the tree, or we can easily remove a given edge from the tree (by removing its two corresponding pairs). The two neighbours *j* and *k* of a vertex *i* are found by searching for the two immediate successors of the pair *(i,-∞)*. The time complexity is, thus, $O(n+m+m \cdot log(m))$, where *n* is the (initial) number of vertices of the graph and *m* is the (initial) number of edges.

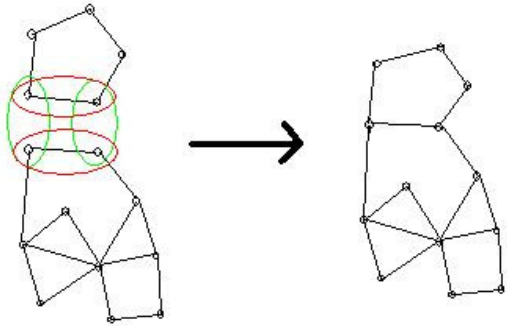

Fig. 4. Attaching a cycle to an edge of a tree of cycles.

### 5. Conclusions and Future Work

This paper presents a novel approach for implementing the Gaussian elimination technique on a 2D SIMD array of processors. The algorithm consists of *2·n-1* iterations and allows row reorderings, by allowing the matrix rows to *slide* past each other. Both the (theoretical) formal proofs of correctness and the (practical) results of the simulation tests validated the concept of the algorithm. Moreover, the paper also introduced several extensions and applications of the Gaussian elimination technique and compared the applications to similar problems in which the Gaussian elimination method was not required for computing the (optimal) solution.

As future work, the algorithm should be implemented on a real architecture with *n·m* processors. Since real-life matrices have sizes which are larger than the number of processors within mainstream parallel computers, a modification of the algorithm in which every processor is responsible for multiple entries of the matrix should be considered. A simple method to achieve this could be to consider that there are *n·m* virtual processors and each real processor simulates several « geographically clustered » virtual processors.

R E F E R E N C E S

[1] *S. F. McGinn, R. E. Shaw,* "Parallel Gaussian Elimination using OpenMP and MPI", in Proceedings of the 16[th] International Symposium on High Performance Computing Systems and Applications, 2002, pp. 169-173.
[2] *P. R. Amestoy, I. S. Duff, J.-Y. L'excellent, X. S. Li,* "Analysis and Comparison of Two General Sparse Solvers for Distributed Memory Computers", in ACM Transactions on Mathematical Software, **vol. 27**, no. 4, 2001, pp. 388-421.
[3] *J. W. Demmel, J. R. Gilbert, X. S. Li,* "An Asynchronous Parallel Supernodal Algorithm for Sparse Gaussian Elimination", in SIAM Journal on Matrix Analysis and Applications, **vol. 20**, no. 4, 1997, pp. 915-952.
[4] *R. Saad,* "An Optimal Schedule for Gaussian Elimination on an MIMD architecture", in Journal of Computational and Applied Mathematics, **vol. 185**, no. 1, 2006, pp. 91-106.
[5] *A. Tiskin,* "Communication-Efficient Parallel Gaussian Elimination", in Lecture Notes in Computer Science, **vol. 2763**, 2003, pp. 369-383.
[6] *B. Hochet, P. Quintin, Y. Robert,* "Systolic Gaussian Elimination Over GF(p) with Partial Pivoting", in IEEE Transactions on Computers, **vol. 38**, no. 9, 1989, pp. 1321-1324.
[7] *V. Cristea*, Algoritmi de prelucrare paralelă, Ed. Matrixrom, Bucureşti, 2002.
[8] *M. Leoncini,* "On the Parallel Complexity of Gaussian Elimination with Pivoting", in Journal of Computer and System Sciences, **vol. 53**, no. 3, 1996, pp. 380-394.
[9] *T. Braunl,* "Parallaxis-III: a structured data-parallel programming language", in Proceedings of the 16[th] International Conference on Algorithms and Architectures for Parallel Processing, 1995, pp. 43-52.
[10] *D. H. Bailey, H. Yozo, X. S. Li, B. Thompson,* "ARPREC: An Arbitrary Precision Computation Package", Technical Report LBNL-53651, Lawrence Berkeley National Laboratory, 2002.
[11] *D. R. Stinson*, Cryptography: Theory and Practice, CRC Press, 2005.
[12] *T. H. Cormen, C. E. Leiserson, R. L. Rivest, C. Stein*, Introduction to Algorithms, 2[nd] edition, MIT Press and McGraw-Hill, 2001.
[13] *V. V. Jinescu, I. Popescu,* "Consideration about High-Pressure Reactors with Cylindrical Symmetry", in U.P.B. Scientific Bulletin, Series D, **vol. 71**, iss. 1, 2009, pp. 45-55.